\documentstyle[12pt,epsfig]{article}
\newcommand{\be}{\begin{equation}}
\newcommand{\ee}{\end{equation}}
\newcommand{\ba}{\begin{eqnarray}}
\newcommand{\ea}{\end{eqnarray}}
\newcommand{\baa}{\begin{eqnarray*}}
\newcommand{\eaa}{\end{eqnarray*}}
\newcommand{\bb}{\begin{thebibliography}}
\newcommand{\eb}{\end{thebibliography}}
\newcommand{\ci}[1]{\cite{#1}}
\newcommand{\bi}[1]{\bibitem{#1}}
\newcommand{\lab}[1]{\label{#1}}

\begin{document}
\sloppy
\thispagestyle{empty}
\mbox{}
\vspace*{\fill}
\begin{center}
{\LARGE {\bf Spin asymmetries of the proton-proton elastic}}\\
\vspace{2mm}
{\LARGE{\bf scattering at HERA-N }}\\

\vspace{2em}
{\large {S.V.Goloskokov, S.P.Kuleshov, O.V.Selyugin}}
\\
\vspace{2em}
{\it  Joint Institut for Nuclear Reserch, \\
 Bogoliubov Laborarory Theoretical Physics, \\
  Dubna, }
 \\
{\it Russia}\\
\end{center}
\vspace*{\fill}
\begin{abstract}
  The spin correlation parameters,
 polarization - $A_{N}$ and double spin
 correlation parameters - $A_{NN}$,
 are calculated in the framework of
 dynamical model (GKS) of the hadron-hadron interection in the whole
 region of small angles and for energy HERA-N $\sqrt{s}=40 GeV$.
 It is shown that the measurements of such spin effects
 at small trunsfer momenta,
 in the region of the dip structure and at $t=-4 GeV^{2}$ have to give
 significant values of such spin correlation parameters.

\end{abstract}
\vspace*{\fill}
\newpage
%
\section{Introduction}
\label{sect1}

    The spin phenomena in diffraction processes
 can give us information about the structure of
 scattering amplitudes of
 hadron-hadron  interaction in the nonperturbative region.

   The measurement of polarization at very small transfer momenta
 up to $ -t = 10^{-4} \ GeV$ can give us the possibility  to find out
 the structure of hadron spin dependent amplitudes.
  In this domain the analyzing power $A_N$
  is determined by the Coulomb-hadron interference effects. As we can
  calculate the Coulomb amplitude very precisely from the theory, we can
  obtain some information about the hadronic non-flip amplitude from this
  quantity.
 If we measure the double spin correlation parameters $A_{NN}$ in that
 domain, we can find out the structure of the hadron double spin--flip
 amplitude \ci{h96n}. Moreover the exact measurement of the point of maximum
 of the coulomb-hadron polarization at small transfer momenta
 gives us the information on how to use the new independent method
 of estimating the total cross section \ci{h96n}.

 The phenomena of interference of the hadronic
 and the coulombic amplitudes may give an important contribution not only
 at very small transfer momenta but also in the range of the diffraction
 minimum \cite{selsp}. So one should know the phase of the interference
 of the coulombic and hadronic amplitude at sufficiently large transfer
 momenta too.

     The majority of theoretical models describe the hadron scattering
 at small angles with the use of
  the eikonal approximation for the scattering amplitude
\ba
 F(s,t) = \frac{1}{2 i} \int_{0}^{\infty} \rho d\rho
         J_0(\rho \Delta)(1-e^{-2i\delta(\rho)}); \lab{feik}
\ea
 where the eikonal phase $  \delta(\rho) $ includes the Coulomb and hadron
 parts and  depends on spin.

  The interaction potential of charged hadrons is
  a sum of Coulomb and nuclear interactions
$$ F^{Born}(s,\Delta,\vec{s}) = F^{born}_{c}(\Delta,\vec{s})
                               +  F^{Born}_{h}(s,\Delta,\vec{s}).$$
 So, after the eikonal summation,
 terms with the Coulomb and nuclear interactions appear.
   The differential cross sections measured in experiment
are described by the square of the scattering amplitude
\begin{eqnarray}
d\sigma /dt =&& \pi \ (F^2_C (t)+ (1 + \rho^{2} (s,t)) \ Im F^2_N(s,t)
                                                             \nonumber \\
 && \mp 2 (\rho (s,t) +\alpha \varphi )) \ F_C (t) \ Im F_N(s,t)),  \label{ds2}
\end{eqnarray}
where $F_{C} = \mp 2 \alpha G^{2}/|t|$ is the Coulomb amplitude;
$\alpha$ is the fine-structure constant  and $G^{2}(t)$ is  the  proton
electromagnetic form factor squared;
$Re\ F_{N}(s,t)$ and $ Im\ F_{N}(s,t)$ are the real and imaginary
parts of the nuclear amplitude;
$\rho(s,t) = Re\ F(s,t) / Im\ F(s,t)$.

  In \ci{selnu} the phase of the Coulomb amplitude
 in the second Born approximation with the form factor
 was calculated in a wide region of transfer momenta.
 It was shown that the behaviors of $\nu$ at non-small $t$ are
 sharply different
 from the behaviors of $\nu$ obtained in \ci{can}.

  In \cite{h96s} was shown
 how we can calculate the total Coulomb--hadron phase that can be used
 in the whole diffraction range of the elastic hadron scattering.

   The obtained eikonal representation for the Coulomb-hadron phase
 is true in a wide region of transfer momenta. If we take the true
 hadron scattering eikonal which describes the experimental differential
 cross sections including the domain of the diffraction dip,
 we can calculate the Coulomb-hadron phase for that region
 of transfer momenta.  This phase will have a real and
 a non-small imaginary part.
    This phase can be very important for the calculation of the spin
 correlation parameters owing to the coulomb-hadron interference effects
 in the domain of the diffraction dip.
  Of course, the largest effects will be in the energy range
  where the diffraction minimum has a sharply defined form and hence
  the real part of the non-flip scattering amplitude will be less.
  So, it will be  in the energy region when
  $20 \leq \sqrt{s} \leq 60 (GeV)$.
 It is clear that for the proton-proton scattering at HERA
  $\sqrt{s} =  40 (GeV)$ we can obtain  significant spin correlation
 effects owing to the Coulomb-hadron interference.

\section{The dynamical model of hadron-hadron
interaction with spin}
\label{sect2}

     In papers \ci{gks}, the dynamical model for a particle interaction
which takes into account the hadron structure at large distances
was developed.
The model is based on the general quantum field theory principles
(analyticity, unitarity, and so on) and takes into account basic information
on the structure of a hadron as a compound system with the central part region
where the valence quarks are concentrated and the long-distance region
where the color-singlet quark-gluon field occurs.
 As a result, the
hadron amplitude can be  represented  as  a  sum  of the central  and
peripheral parts of the interaction:
\ba
T(s,t) \propto T_{c}(s,t) + T_{p}(s,t).  \lab{tt}
\ea
 where $T_{c}(s,t)$   describes   the
interaction between the central parts of hadrons.
At high energies it is determined by the spinless pomeron exchange.
The quantity  $T_{p}(s,t)$  is
a sum of triangle diagrams corresponding to
the interactions of the central part of one hadron with  the  meson
cloud of the other. The meson - nucleon interaction leads to the spin flip
effects at the pomeron-hadron vertex.

          The contribution of these triangle diagrams
to the scattering amplitude  with  $N(\Delta $-isobar)  in  the
intermediate state looks  as follows \ci{zpc}:

\ba
T ^{\lambda _{1}\lambda _{2}}_{N(\Delta )}(s,t) =
{g^{2}_{\pi NN(\Delta )} \over i(2\pi )^{4}}
  \int  d^{4} q T_{\pi N} (s^{\prime}, t) \varphi_{N (\Delta )}
[(k-q),q^{2}] \varphi_{N(\Delta )} [(p-q),q^{2}]    \nonumber \\
\times \frac{\Gamma^{\lambda_{1}\lambda_{2}}(q,p,k,)}
{
[q^{2} - M^{2}_{N (\Delta) } + i \epsilon ] [(k-q)^{2} - \mu^{2} +
 i \epsilon ]  [(p-q)^{2} - \mu^{2} + i \epsilon ]
}.   \lab{tint}
\ea
Here $\lambda_{1},\lambda_{2}$  are  helicities  of  nucleons;
$T_{\pi N}$  is  the
$\pi N$-scattering amplitude; $\Gamma$  is a matrix element of the numerator
of the diagram ; $\varphi $ are vertex functions chosen
 in the dipole form  with
the parameters $\beta _{N(\Delta )}$:
\ba
\varphi _{N(\Delta )}(l^{2},q^{2}\propto  M^{2}_{N(\Delta )})
  = {b^{4}_{N(\Delta )}\over
      (b^{2}_{N(\Delta )}- l^{2})^{2}}.
   \lab{fi}
\ea

       For a standard form of the pomeron contribution to the
meson-nucleon scattering amplitude
$$
    T_{\pi N} (s,t) = i \beta^{\pi}(t) \cdot \beta^{N}(t) s^{\alpha(t)}
$$
we can write the integral (\ref{tint}) in the form:
$$
    T^{\lambda_1 \lambda_2}_{N (\Delta)}(s,t) =
  i \beta^{N(\lambda_1 \lambda_2)}(t) \cdot \beta^{N}(t) s^{\alpha(t)},
$$
where  $\beta^{N(\lambda_1 \lambda_2)}(t)$ is the $N$-nucleon or
$\Delta_{33}$-isobar contribution to a spin-dependent nucleon-pomeron
vertex function.
     The  peripheral
contribution calculated in the model leads to the  spin  effects
in the Born term of the scattering amplitude which do not disappear
with growing energy.
 Summation of rescatterings in s-channel has been performed with the help
of the quasipotential equation. The total amplitude has an eikonal form.
The explicit forms of helicity amplitudes and parameters
obtained can be found in \ci{zpc}.

        The model with  the $N $  and  $\Delta $
contribution  provides  a self-consistent  picture  of
the differential cross sections and spin phenomena
of different hadron processes  at  high  energies.
Really, the parameters in the  amplitude  determined from one
reaction, for example, elastic $pp$-scattering, allow one to  obtain
a wide range of results for elastic
meson-nucleon scattering and charge-exchange reaction
 $\pi^{-} p \rightarrow  \pi^{0} n$
 at high energies.

\section{ The model results at HERA-N energy}
\label{sect3}

 In Fig.1, shown are the model calculations for the analyzing power
 of the proton-proton scattering
 at $sqrt{s} = 23.4 \ GeV $ where we have the experimental data.
\begin{center}
\mbox{\epsfig{file=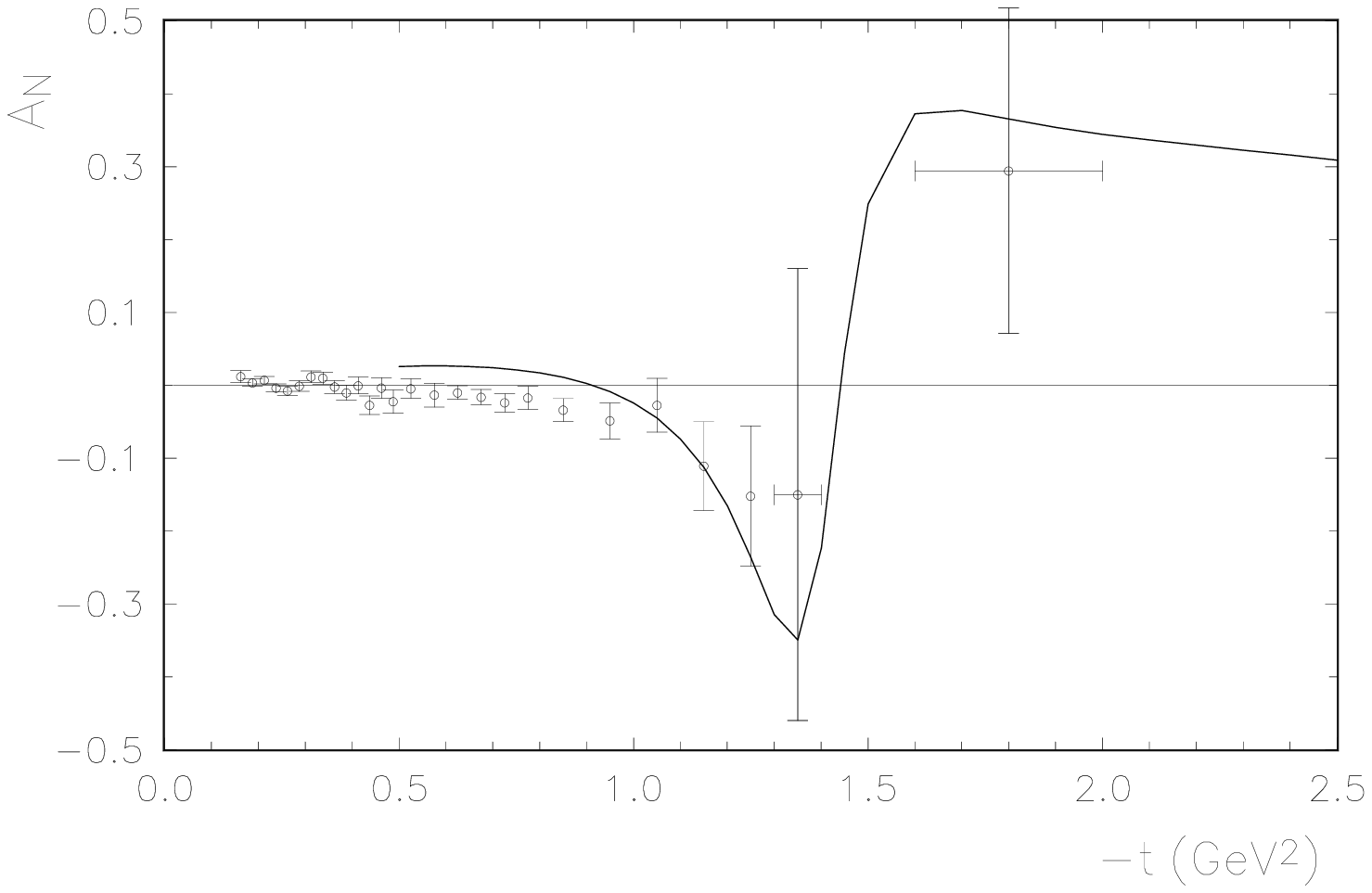, height=8cm,width=11.5cm}}
\noindent
\small
\begin{minipage}{9cm}
\vspace{7mm}
{\bf Fig.~1:}
 Polarization of the $ p p$ - scattering  calculated in the model GKS
 at $\sqrt{s}=23.4 \ GeV$
and the experimental points \ci{kl}.
\end{minipage}
\end{center}
   It is clear that the model reproduces the experimental
 data well and we can think that the prediction of the model at
 HERA-N  energy will be also sufficient good.
  This prediction for the analyzing power and the double spin
correlation parameter  $A_{NN}$   is shown in Figs. 2 and 3.
     The model predicts that at superhigh energies  the  polarization
effects of particles and antiparticles are the same.
After $\sqrt{s}=50 \ GeV$, the analyzing power
decreases very slowly and has a very determined form.
It is small at small transfer momenta, before the diffraction peak,
and has a narrow sufficiently large negative peak in the range
of the diffraction minimum, see Fig. 2.
\linebreak
\begin{center}
\begin{tabular}{cc}
\mbox{\epsfig{file=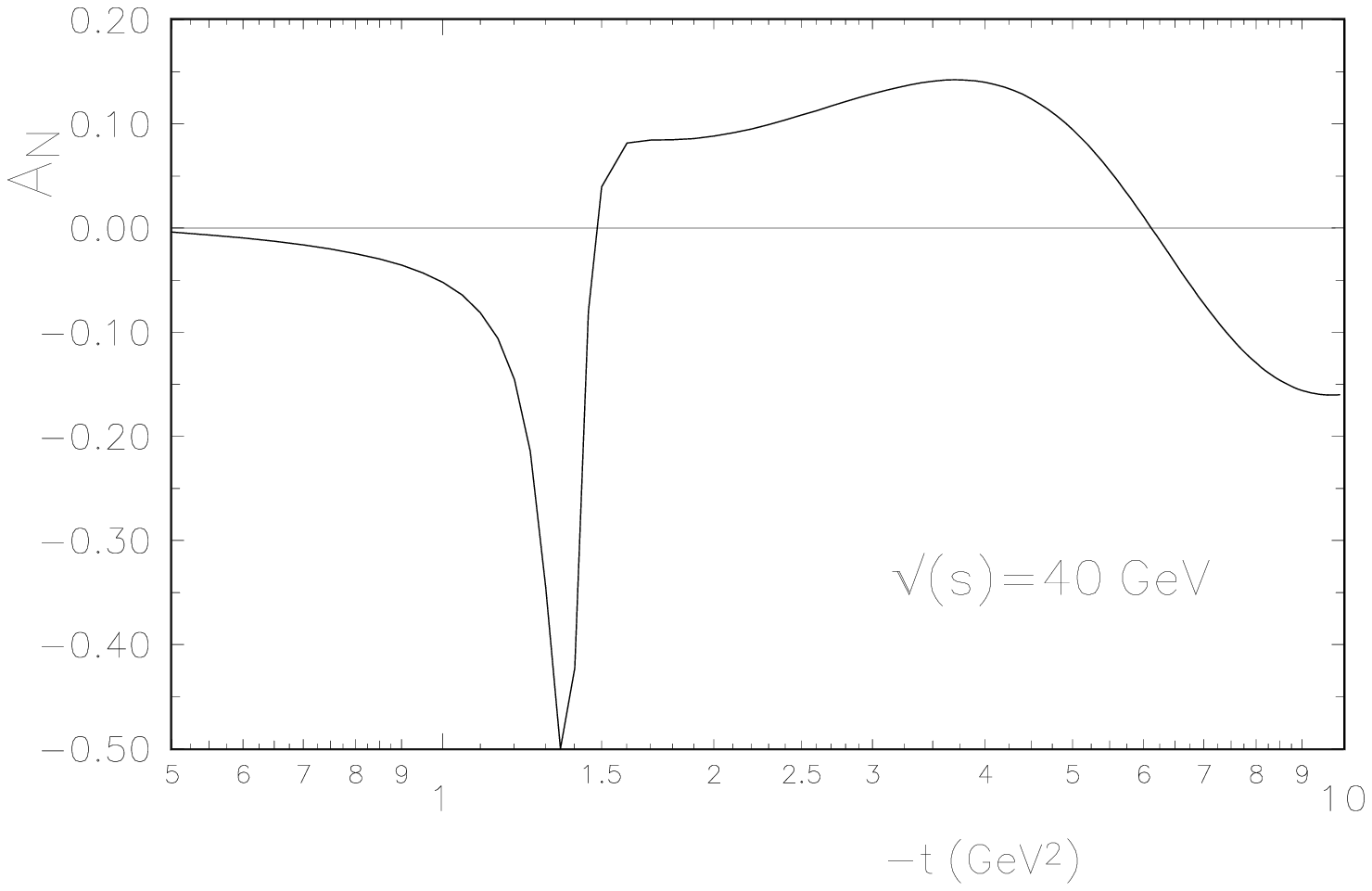,height=7.5cm,width=7.5cm}}
\vspace{2mm}
\noindent
\small
&
\mbox{\epsfig{file=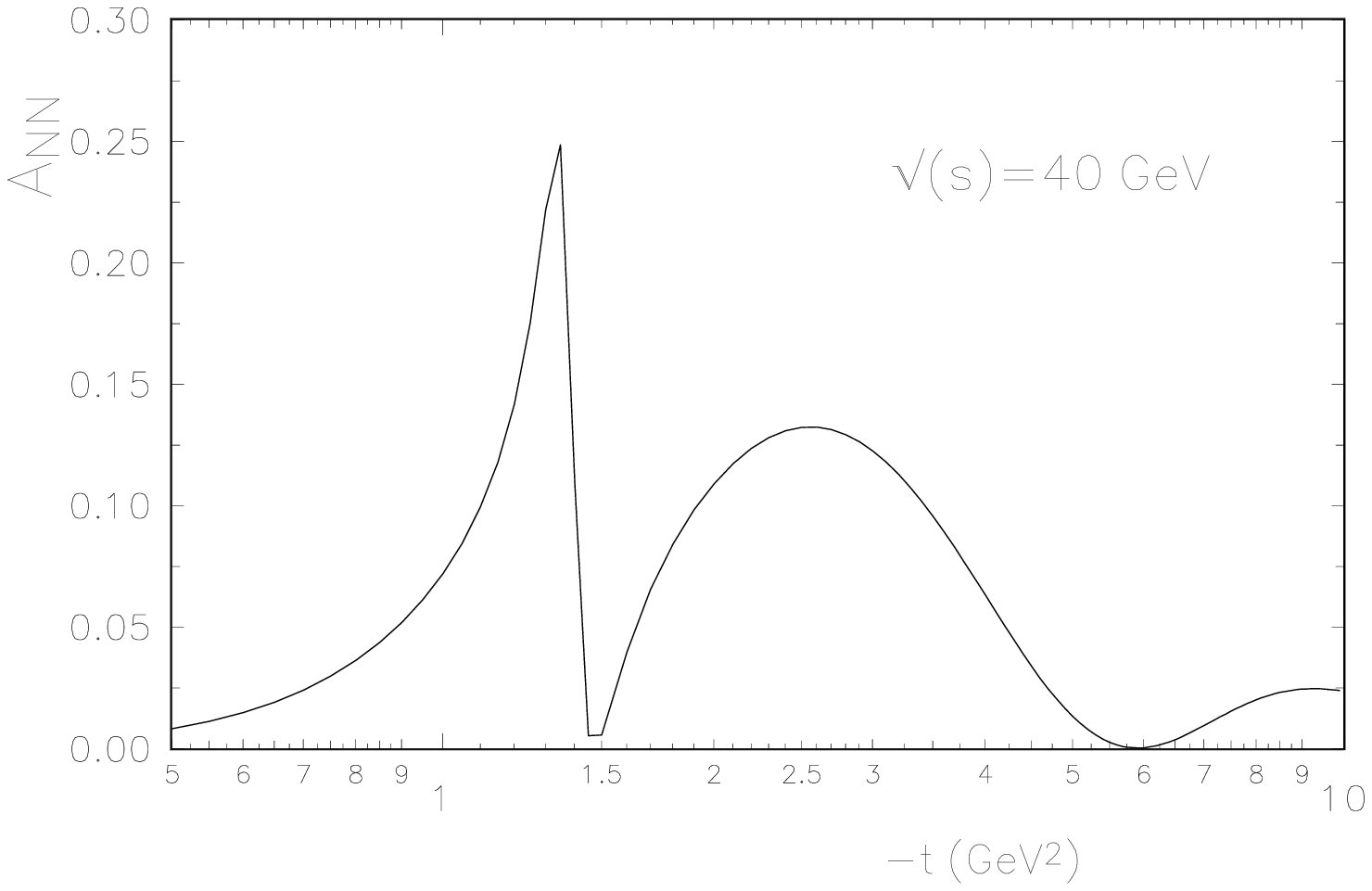,height=7.5cm,width=7.5cm}}
\vspace{2mm}
\noindent
\small
\\
\begin{minipage}{6cm}
{\sf Figure~2:}
 Analyzing power $A_N$ of the $ p p$ - scattering  calculated in the
 model GKS
at $\sqrt{s}=40  \  GeV$
\end{minipage}
 \normalsize
&
\begin{minipage}{6cm}
{\sf Figure~3:}
 Double spin correlation parameter $A_{NN}$
of the $ p p$ - scattering  calculated in the model GKS
at $\sqrt{s}=40  \  GeV$
\end{minipage}
\normalsize
\end{tabular}
\end{center}
\vspace{0.5cm}

Behind the diffraction minimum $A_N$
changes its sign and has the bump up to $|t|=6 GeV^2$.
The position of maximum of this bump slowly changes towards larger $|t|$
and its magnitude somewhat changes around $10\%$;
near $|t|=3 GeV^2$ the magnitude is practically
 constant, $\simeq 8\%$.
      The behavior of the spin correlation parameter $A_{NN}$ is  shown
in fig.3.
 As is seen, the value of $A_{NN}$  becomes  maximum  in  the
range of the  diffraction  minimum,  as  in  the  case  of
the polarization.
The magnitude of $A_{NN}$ becomes sufficiently large with the growth of $|t|$.
The reason is that in this work we have used
the strong form factors for the vertices $\pi N N $  and $\pi N \Delta $.
The form of the spin-flip amplitude is determined in
the model up to $\mid t \mid  \simeq  2$ GeV$^{2}$, hence,  we  can
expect  an  adequate
description of the experimental data up to
$\mid t\mid  \simeq  3 \div  5.0$  GeV$^{2}$.

 Note that  the
polarizations of $pp$-  and $p\bar{p}$-  scattering  will  coincide  above
the energies $\sqrt{s}> 30 GeV$.

    Thus, the dynamical model considered, which takes into account
 the $N$ and $\Delta$ contribution, leads
 to a lot of predictions concerning
 the behavior of spin correlation parameters at high energies.
 In that model, the effects of large  distances
 determined  by  the  meson  cloud  of  hadrons  give  a   dominant
 contribution to the spin-flip amplitudes  of  different  exclusive
 processes at high energies and fixed transfer momenta.
 Note that the results on the spin effects obtained here differ from
 the predictions of other models \ci{BS} at an energy above
 $\sqrt{s} \geq 30 GeV$.
 And the examination of these results gives new information about the
 hadron interaction at large distances.

\section{Conclusion}
\label{sect4}
    In the HERA-N experimental energy region  {($\sqrt{s}=40 GeV$ ) we have
  proton-proton elastic scattering with a very deep
 first diffraction minimum.
 It is connected with the energy dependence of the real part of the spin
 non-flip amplitude. At this energy the real part  is small
 and  gives a small contribution at the point of the diffraction minimum.
 The relation
 of the real to imaginary parts of the spin-non flip amplitude
 $\rho$ equals $ 0.06$  at small transfer momenta.
 In this case we can obtain the spin effects larger
 in the diffraction dip domain
 than in other
 high energy experiments at $\sqrt{s} \geq 50 \ GeV$ where  $\rho \geq 0.095$.
 At the energy $\sqrt{s} = 40 \ GeV$
 we can more exactly explore the spin effect of the
 coulomb-hadron interference, as the uncertainty of the  real part of
 the spin-non flip amplitude gives a small distortion and the
  contribution of the hadron spin--flip amplitude will be large
 and defined fluently.
  So, the measurement of  spin effects in the elastic scattering
 at small angles at HERA-N is not the competition with  other
 spin programs at RHIC and LHC but is the cooperation
 which gives the first and very important step
 in explore of the super-high energy spin physics.

\section*{Acknowledgement}
The   authors   express   their   gratitude   to
 N.Akchurin, A.V.Efremov, M. Islam, W.D.Nowak, S.B.Nurushev,
 for the discussion of the problems of the measure of the spin effects
 at RHIC and HERA energies.


\end{document}